\documentclass[english]{iopart}
\usepackage[T1]{fontenc}
\usepackage[latin9]{inputenc}
\usepackage{textcomp}
\usepackage{amstext}
\usepackage{amssymb}
\usepackage{esint}

\makeatletter

\newcommand{\noun}[1]{\textsc{#1}}
\newcommand{\lyxmathsym}[1]{\ifmmode\begingroup\def\b@ld{bold}
  \text{\ifx\math@version\b@ld\bfseries\fi#1}\endgroup\else#1\fi}

\usepackage{iopams}
\usepackage{setstack}

\makeatother

\usepackage{babel}
\begin{document}

\title{Inflation and topological phase transition driven by exotic smoothness}

\author{Torsten Asselmeyer-Maluga}

\ead{torsten.asselmeyer-maluga@dlr.de}

\address{German Aerospace Center (DLR), Berlin, Germany}

\author{Jerzy Kr\'ol}

\ead{iriking@wp.pl}

\address{University of Silesia, Institute of Physics, Katowice, Poland}
\begin{abstract}
In this paper we will discuss a model which describes the cause of
inflation by a topological transition. The guiding principle is the
choice of an exotic smoothness structure for the space-time. Here
we consider a space-time with topology $S^{3}\times\mathbb{R}$. In
case of an exotic $S^{3}\times\mathbb{R}$, there is a change in the
spatial topology from a 3-sphere to a homology 3-sphere which can
carry a hyperbolic structure. From the physical point of view, we
will discuss the path integral for the Einstein-Hilbert action with
respect to a decomposition of the space-time. The inclusion of the
boundary terms produces fermionic contributions to the partition function.
The expectation value of an area (with respect to some surface) shows
an exponential increase, i.e. we obtain inflationary behavior. We
will calculate the amount of this increase to be a topological invariant.
Then we will describe this transition by an effective model, the Starobinski
or $R^{2}$ model which is consistent with the current measurement
of the Planck satellite. The spectral index and other observables
are also calculated.
\end{abstract}
\maketitle

\section{Introduction}

General relativity (GR) has changed our understanding of space-time.
In parallel, the appearance of quantum field theory (QFT) has modified
our view of particles, fields and the measurement process. The usual
approach for the unification of QFT and GR, to a quantum gravity,
starts with a proposal to quantize GR and its underlying structure,
space-time. There is a unique opinion in the community about the relation
between geometry and quantum theory: The geometry as used in GR is
classical and should emerge from a quantum gravity in the limit (Planck's
constant tends to zero). Most theories went a step further and try
to get a space-time from quantum theory. Then, the model of a smooth
manifold is not suitable to describe quantum gravity. But, there is
no sign for a discrete space-time structure or higher dimensions in
current experiments. Hence, quantum gravity based on the concept of
a smooth manifold should also able to explain the current problems
in the standard cosmological model ($\Lambda$CDM) like the appearance
of dark energy/matter, or the correct form of inflation etc. But before
we are going in this direction we will motivate the usage of the smooth
manifold as our basic concept.

When Einstein developed GR, his opinion about the importance of general
covariance changed over the years. In 1914, he wrote a joint paper
with Grossmann. There, he rejected general covariance by the now famous
hole argument. But after a painful year, he again considered general
covariance now with the insight that there is no meaning in referring
to the \emph{space-time point A} or the \emph{event A}, without further
specifications. Therefore the measurement of a point without a detailed
specification of the whole measurement process is meaningless in GR.
The reason is simply the diffeomorphism-invariance of GR which has
tremendous consequences. Furthermore, GR do not depend on the topology
of space-time. All restrictions on the topology of the space-time
were formulated using additional physical conditions like causality
(see \cite{HawEll:94}). This ambiguity increases in the 80's when
the first examples of exotic smoothness structures in dimension 4
were found. The (smooth) atlas of a smooth 4-manifold $M$ is called
the smoothness structure (unique up to diffeomorphisms). One would
expect that there is only one smooth atlas for $M$, all other possibilities
can be transformed into each other by a diffeomorphism. But in contrast,
the deep results of Freedman \cite{Fre:82} on the topology of 4-manifolds
combined with Donaldson's work \cite{Don:83} gave the first examples
of non-diffeomorphic smoothness structures on 4-manifolds including
the well-known $\mathbb{R}^{4}$. Much of the motivation can be found
in the FQXI essay \cite{AsselmeyerFQXI-Essay2012}. Here we will discuss
another property of the exotic smoothness structure: its quantum geometry
in the path integral.

Diffeomorphism invariance is the most important property of the Einstein-Hilbert
action with far reaching consequences\cite{Pfeiffer2004}. One of
our results is a close relation between geometry and foliation to
exotic smoothness \cite{AsselmeyerKrol2009,AsselmeyerKrol2011d}.
In the particular example of the exotic $\mathbb{R}^{4}$, we discussed
the exotic smoothness structures as a manifestation of quantum gravity
(by using string theory \cite{AsselmeyerKrol2011,AsselmeyerKrol2011b}).
This exotic $\mathbb{R}^{4}$ has some interesting properties as first
noted by Brans \cite{Bra:94a,Bra:94b}. More importantly as shown
by S{\l}adkowski \cite{Sladkowski2001}, the exotic $\mathbb{R}^{4}$
has a non-trivial curvature in contrast to the flat standard $\mathbb{R}^{4}$.
It was the first result that an exotic $\mathbb{R}^{4}$ can be seen
as a source of gravity (or it must contain sources of gravity). S{\l}adkowski
\cite{Sla:96,Sla:96b,Sla:96c} went further and showed a relation
to particle physics also related to quantum gravity. But why there
is a relation to quantum gravity? In \cite{AsselmeyerKrol2011c} we
presented the first idea to understand this relation which was further
extended in \cite{AsselmeyerKrol2013}. An exotic 4-manifold like
$S^{3}\times\mathbb{R}$ is also characterized by the property that
there is no smoothly embedded 3-sphere but a topological embedded
one. This topological $S^{3}$ is wildly embedded, i.e. the image
of the embedding must be triangulated by an infinite polyhedron. In
\cite{AsselmeyerKrol2013}, we proved that the (deformation) quantization
of a usual (or tame) embedding is a wild embedding which can be seen
as a quantum state. But then any exotic 4-manifold can be interpreted
as a quantum state of the 4-manifold with standard smoothness structure.
From this point of view, the calculation of the path integral in quantum
gravity has to include the exotic smoothness structures. Usually it
is hopeless to make these calculations. But by using the close relation
of exotic smoothness to hyperbolic geometry, one has a chance to calculate
geometric expressions like the expectation value of the surface area.
In this paper we will show that this expectation value has an inflationary
behavior, i.e. the area grows exponentially (along the time axis).
Therefore quantum gravity (in the sense of exotic smoothness) can
be the root of inflation.

\section{space-time and smoothness}

From the mathematical point of view, the space-time is a smooth 4-manifold
endowed with a (smooth) metric as basic variable for general relativity.
The existence question for Lorentz structure and causality problems
(see Hawking and Ellis \cite{HawEll:94}) give further restrictions
on the 4-manifold: causality implies non-compactness, Lorentz structure
needs a non-vanishing normal vector field. Both concepts can be combined
in the concept of a global hyperbolic 4-manifold $M$ having a Cauchy
surface $\mathcal{S}$ so that $M=\mathcal{S}\times\mathbb{R}$. 

All these restrictions on the representation of space-time by the
manifold concept are clearly motivated by physical questions. Among
these properties there is one distinguished element: the smoothness.
Usually one starts with a topological 4-manifold $M$ and introduces
structures on them. Then one has the following ladder of possible
structures:
\begin{eqnarray*}
\mbox{Topology}\to & \mbox{\mbox{piecewise-linear(PL)}}\to & \mbox{Smoothness}\to\\
\qquad\to & \mbox{bundles, Lorentz, Spin etc.}\to & \mbox{metric, geometry,...}
\end{eqnarray*}
We do not want to discuss the first transition, i.e. the existence
of a triangulation on a topological manifold. But we remark that the
existence of a PL structure implies uniquely a smoothness structure
in all dimensions smaller than 7 \cite{KirSie:77}. Here we have to
consider the following steps to define a space-time:
\begin{enumerate}
\item Fix a topology for the space-time $M$.
\item Fix a smoothness structure, i.e. a maximal differentiable atlas $\mathcal{A}$.
\item Fix a smooth metric or get one by solving the Einstein equation.
\end{enumerate}
The choice of a topology never fixes the space-time uniquely, i.e.
there are two space-times with the same topology which are not diffeomorphic.
The main idea of the paper is the introduction of exotic smoothness
structures into space-time. If two manifolds are homeomorphic but
non-diffeomorphic, they are \textbf{exotic} to each other. The smoothness
structure is called an \textbf{exotic smoothness structure}.

In dimension four there are many examples of compact 4-manifolds with
countable infinite non-diffeomorphic smoothness structures and many
examples of non-compact 4-manifolds with uncountable infinite many
non-diffeomorphic smoothness structures. But in contrast, the number
of non-diffeomorphic smoothness structures is finite for any other
dimension \cite{KirSie:77}. As an example, we will consider the space-time
$S^{3}\times\mathbb{R}$ having uncountable many non-diffeomorphic
smoothness structures in the following.

\section{The path integral in exotic $S^{3}\times\mathbb{R}$}

For simplicity, we consider general relativity without matter (using
the notation of topological QFT). Space-time is a smooth oriented
4-manifold $M$ which is non-compact and without boundary. From the
formal point of view (no divergences of the metric) one is able to
define a boundary $\partial M$ at infinity. The classical theory
is the study of the existence and uniqueness of (smooth) metric tensors
$g$ on $M$ that satisfy the Einstein equations subject to suitable
boundary conditions. In the first order Hilbert\textendash{}Palatini
formulation, one specifies an $SO(1,3)$-connection $A$ together
with a cotetrad field $e$ rather than a metric tensor. Fixing $A|_{\partial M}$
at the boundary, one can derive first order field equations in the
interior (now called \emph{bulk}) which are equivalent to the Einstein
equations provided that the cotetrad is non-degenerate. The theory
is invariant under space-time diffeomorphisms $M\to M$. In the particular
case of the space-time $M=S^{3}\times\mathbb{R}$ (topologically),
we have to consider a smooth 4-manifolds $M_{i,f}$ as parts of $M$
whose boundary $\partial M_{i,f}=\Sigma_{i}\sqcup\Sigma_{f}$ is the
disjoint union of two smooth 3-manifolds $\Sigma_{i}$ and $\Sigma_{f}$
to which we associate Hilbert spaces $\mathcal{H}_{j}$ of 3-geometries,
$j=i,f$. These contain suitable wave functionals of connections $A|_{\Sigma_{j}}$
. We denote the connection eigenstates by $|A|_{\Sigma_{j}}\rangle$.
The path integral,
\begin{equation}
\langle A|_{\Sigma_{f}}|T_{M}|A|_{\Sigma_{i}}\rangle=\intop_{A|\partial M_{i,f}}DA\, De\,\exp\left(\frac{i}{\hbar}S_{EH}[e,A,M_{i,f}]\right)\label{eq:path-integral}
\end{equation}
is the sum over all connections $A$ matching $A|_{\partial M_{i,f}}$,
and over all $e$. It yields the matrix elements of a linear map $T_{M}:\mathcal{H}_{i}\to\mathcal{H}_{f}$
between states of 3-geometry. Our basic gravitational variables will
be cotetrad $e_{a}^{I}$ and connection $A_{a}^{IJ}$ on space-time
$M$ with the index $a$ to present it as 1-forms and the indices
$I,J$ for an internal vector space $V$ (used for the representation
of the symmetry group). Cotetrads $e$ are \textquoteleft{}square-roots\textquoteright{}
of metrics and the transition from metrics to tetrads is motivated
by the fact that tetrads are essential if one is to introduce spinorial
matter. $e_{a}^{I}$ is an isomorphism between the tangent space $T_{p}(M)$
at any point $p$ and a fixed internal vector space $V$ equipped
with a metric $\eta_{IJ}$ so that $g_{ab}=e_{a}^{I}e_{b}^{J}\eta_{IJ}$.
Here we used the action
\begin{equation}
S_{EH}[e,A,M_{i,f},\partial M_{i,f}]=\intop_{M_{i,f}}\epsilon_{IJKL}(e^{I}\wedge e^{J}\wedge\left(dA+A\wedge A\right)^{KL})+\intop_{\partial M_{i,f}}\epsilon_{IJKL}(e^{I}\wedge e^{J}\wedge A^{KL})\label{eq:action-with-boundary}
\end{equation}
in the notation of \cite{Ashtekar08,Ashtekar08a}. Here the boundary
term $\epsilon_{IJKL}(e^{I}\wedge e^{J}\wedge A^{KL})$ is equal to
twice the trace over the extrinsic curvature (or the mean curvature).
For fixed boundary data, (\ref{eq:path-integral}) is a diffeomorphism
invariant in the bulk. If $\Sigma_{i}=\Sigma_{f}$ are diffeomorphic,
we can identify $\Sigma=\Sigma_{i}=\Sigma_{f}$ and $\mathcal{H}=\mathcal{H}_{i}=\mathcal{H}_{f}$
i.e. we close the manifold $M_{i,f}$ by identifying the two boundaries
to get the closed 4-manifold $M'$. Provided that the trace over $\mathcal{H}$
can be defined, the partition function, 
\begin{equation}
Z(M')=tr_{\mathcal{H}}T_{M}=\int DA\, De\,\exp\left(\frac{i}{\hbar}S_{EH}[e,A,M,\partial M]\right)\label{eq:path-integral-1}
\end{equation}
where the integral is now unrestricted, is a dimensionless number
which depends only on the diffeomorphism class of the smooth manifold
$M\lyxmathsym{\textasciiacute}$. In case of the manifold $M_{i,f}$,
the path integral (as transition amplitude) $\langle A|_{\Sigma_{f}}|T_{M}|A|_{\Sigma_{i}}\rangle$
is the diffeomorphism class of the smooth manifold relative to the
boundary. But the diffeomorphism class of the boundary is unique and
the value of the path integral depends on the topology of the boundary
as well on the diffeomorphism class of the interior of $M_{i,f}$.
Therefore we will shortly write
\[
\langle\Sigma_{f}|T_{M}|\Sigma_{i}\rangle=\langle A|_{\Sigma_{f}}|T_{M}|A|_{\Sigma_{i}}\rangle
\]
and consider the sum of manifolds like $M_{i,h}=M_{i,f}\cup_{\Sigma_{f}}M_{f,h}$
with the amplitudes
\begin{equation}
\langle\Sigma_{h}|T_{M}|\Sigma_{i}\rangle=\sum_{A|\Sigma_{f}}\langle\Sigma_{h}|T_{M}|\Sigma_{f}\rangle\langle\Sigma_{f}|T_{M}|\Sigma_{i}\rangle\label{eq:decomposition-amplitudes}
\end{equation}
where we sum (or integrate) over the connections and frames on $\Sigma_{h}$
(see \cite{hawkingpathintegral1979}). Then the boundary term
\[
S_{\partial}[\Sigma_{f}]=\intop_{\Sigma_{f}}\epsilon_{IJKL}(e^{I}\wedge e^{J}\wedge A^{KL})=\intop_{\Sigma_{f}}H\sqrt{h}d^{3}x
\]
is needed where $H$ is the mean curvature of $\Sigma_{f}$ corresponding
to the metric $h$ at $\Sigma_{f}$ (as restriction of the 4-metric).
Therefore we have to divide the path integration into two parts: the
contribution by the boundary (boundary integration) and the contribution
by the interior (bulk integration).

\subsection{Boundary integration}

The boundary $\Sigma$ of a 4-manifold $M$ can be understood as embedding
(or at least as immersion). Let $\iota:\Sigma\hookrightarrow M$ be
an immersion of the 3-manifold $\Sigma$ into the 4-manifold $M$
with the normal vector $\vec{N}$. The spin bundle $S_{M}$ of the
4-manifold splits into two sub-bundles $S_{M}^{\pm}$ where one subbundle,
say $S_{M}^{+},$ can be related to the spin bundle $S_{\Sigma}$
of the 3-manifold. Then the spin bundles are related by $S_{\Sigma}=\iota^{*}S_{M}^{+}$
with the same relation $\phi=\iota_{*}\Phi$ for the spinors ($\phi\in\Gamma(S_{\Sigma})$
and $\Phi\in\Gamma(S_{M}^{+})$). Let $\nabla_{X}^{M},\nabla_{X}^{\Sigma}$
be the covariant derivatives in the spin bundles along a vector field
$X$ as section of the bundle $T\Sigma$. Then we have the formula
\begin{equation}
\nabla_{X}^{M}(\Phi)=\nabla_{X}^{\Sigma}\phi-\frac{1}{2}(\nabla_{X}\vec{N})\cdot\vec{N}\cdot\phi\label{eq:covariant-derivative-immersion}
\end{equation}
with the obvious embedding $\phi\mapsto\left(\begin{array}{c}
\phi\\
0
\end{array}\right)=\Phi$ of the spinor spaces. The expression $\nabla_{X}\vec{N}$ is the
second fundamental form of the immersion where the trace $tr(\nabla_{X}\vec{N})=2H$
is related to the mean curvature $H$. Then from (\ref{eq:covariant-derivative-immersion})
one obtains a similar relation between the corresponding Dirac operators
\begin{equation}
D^{M}\Phi=D^{3D}\phi-H\phi\label{eq:relation-Dirac-3D-4D}
\end{equation}
with the Dirac operator $D^{3D}$ of the 3-manifold $\Sigma$. Near
the boundary $\Sigma$, the 4-manifolds looks like $\Sigma\times[0,1]$
and a spinor $\Phi$ on this 4-manifold is a parallel spinor and has
to fulfill the equation 
\begin{equation}
D^{M}\Phi=0\label{eq:Dirac-equation-4D}
\end{equation}
i.e. $\phi$ yields the eigenvalue equation
\begin{equation}
D^{3D}\phi=H\phi\label{eq:Dirac-equation-3D}
\end{equation}
with the mean curvature $H$ of the embedding $\iota$ as eigenvalue.
See our previous work \cite{AsselmeyerRose2012} for more details.

Now we will use this theory to get rid of the boundary integration.
At first we will discuss the deformation of a immersion using a diffeomorphism.
Let $I:\Sigma\hookrightarrow M$ be an immersion of $\Sigma$ (3-manifold)
into $M$ (4-manifold). A deformation of an immersion $I':\Sigma'\hookrightarrow M'$
are diffeomorphisms $f:M\to M'$ and $g:\Sigma\to\Sigma'$ of $M$
and $\Sigma$, respectively, so that 
\[
f\circ I=I'\circ g\,.
\]
One of the diffeomorphism (say $f$) can be absorbed into the definition
of the immersion and we are left with one diffeomorphism $g\in Diff(\Sigma)$
to define the deformation of the immersion $I$. But as stated above,
the immersion is directly given by an integral over the spinor $\phi$
on $\Sigma$ fulfilling the Dirac equation (\ref{eq:Dirac-equation-3D}).
Therefore we have to discuss the action of the diffeomorphism group
$Diff(\Sigma)$ on the Hilbert space of $L^{2}-$spinors fulfilling
the Dirac equation. This case was considered in the literature \cite{SpinorsDiffeom2013}.
The spinor space $S_{g,\sigma}(\Sigma)$ on $\Sigma$ depends on two
ingredients: a (Riemannian) metric $g$ and a spin structure $\sigma$
(labeled by the number of elements in $H^{1}(\Sigma,\mathbb{Z}_{2})$).
Let us consider the group of orientation-preserving diffeomorphism
$Diff^{+}(\Sigma)$ acting on $g$ (by pullback $f^{*}g$) and on
$\sigma$ (by a suitable defined pullback $f^{*}\sigma$). The Hilbert
space of $L^{2}-$spinors of $S_{g,\sigma}(\Sigma)$ is denoted by
$H_{g,\sigma}$. Then according to \cite{SpinorsDiffeom2013}, any
$f\in Diff^{+}(\Sigma)$ leads in exactly two ways to a unitary operator
$U$ from $H_{g,\sigma}$ to $H_{f^{*}g,f^{*}\sigma}$. The (canonically)
defined Dirac operator is equivariant with respect to the action of
$U$ and the spectrum is invariant under (orientation-preserving)
diffeomorphisms. In particular we obtain for the boundary term
\[
S_{\partial}[\Sigma_{f},h]=\intop_{\Sigma_{f}}H\sqrt{h}d^{3}x=\intop_{\Sigma_{f}}\overline{\phi}\, D^{3D}\phi d^{3}x
\]
with $|\phi|^{2}=const.$(see \cite{Friedrich1998}). But then we
can change the integration process from the integration over the metric
class $h$ on the 3-manifold $\Sigma_{f}$ with mean curvature to
an integration over the spinor $\phi$ on $\Sigma_{f}$. Then we obtain
\begin{eqnarray}
Z(\Sigma_{f}) & = & \intop Dh\,\exp\left(\frac{i}{\hbar}S[\Sigma_{f},h]\right)=\int D\phi D\bar{\phi}\,\exp\left(\frac{i}{\hbar}\intop_{\Sigma_{f}}\overline{\phi}\, D^{3D}\phi d^{3}x\right)\nonumber \\
 & = & \sqrt{\det\left(D^{3D}D^{*3D}\right)}e^{i\pi\eta(\Sigma_{f})/2}\label{eq:path-integral-boundary}
\end{eqnarray}
where $\eta(\Sigma_{f})$ is the Eta invariant of the Dirac operator
at the 3-manifold $\Sigma_{f}$ (here we use a result of Witten see
\cite{Wit:85}).

From the physical point of view, we obtain fermions at the boundary.
The additional term with the Eta invariant reflects also an important
fact. The state space of general relativity is the space of the (Lorentzian)
metric tensor up to the group of coordinate transformations. This
group of coordinate transformations is not the full diffeomorphism
group, it is only one connected component of the diffeomorphism. That
is the group of diffeomorphisms connected to the identity. In addition,
there is also the (discrete) group of global diffeomorphisms which
is in our case detected by the Eta invariant. For 3-manifolds there
is a deep relation to the Chern-Simons invariant \cite{Yoshida:1985}
which will be further studied at our forthcoming work.

\subsection{Bulk integration}

Now we will discuss the path integral of the action 
\[
S_{EH}[e,A,M]=\intop_{M}\epsilon_{IJKL}(e^{I}\wedge e^{J}\wedge\left(dA+A\wedge A\right)^{KL})
\]
in the interior of the 4-manifold $M$. The contribution of the boundary
was calculated in the previous subsection. In the (formal) path integral
(\ref{eq:path-integral}) we will ignore all problems (ill-definiteness,
singularities etc.) of the path integral approach. Next we have to
discuss the measure $De$ of the path integral. Currently there is
no rigorous definition of this measure and as usual we assume a product
measure. 

Then we have two possible parts which are more or less independent
from each other:
\begin{enumerate}
\item integration $De_{G}$ over geometries 
\item integration $De_{DS}$ over different differential structures parametrized
by some structure (see below).
\end{enumerate}
Now we have to consider the following path integral

\[
Z(M)=\intop_{Diff\, structures}De_{DS}\left(\intop_{Geometries}De_{G}\:\exp\left(\frac{i}{\hbar}S_{EH}[e,M]\right)\right)
\]
and we have to calculate the influence of the differential structures
first. At this level we need an example, an exotic $S^{3}\times\mathbb{R}$.

\subsection{Constructing exotic $S^{3}\times\mathbb{R}$}

In \cite{Fre:79}, Freedman constructed the first example of an exotic
$S^{3}\times\mathbb{R}$ of special type. There are also uncountable
many different exotic $\mathbb{R}^{4}$ having an end homeomorphic
to $S^{3}\times\mathbb{R}$ but not diffeomorphic to it. But Freedmans
first example is not of this type (as an end of an exotic $\mathbb{R}^{4}$).
Therefore to get an infinite number of different exotic $S^{3}\times\mathbb{R}$
one has to see $S^{3}\times\mathbb{R}$ as an end of $\mathbb{R}^{4}$
also expressible as complement $\mathbb{R}^{4}\setminus D^{4}$ of
the 4-disk. A second possibility is the usage of the end-sum technique
of Gompf, so that the standard $S^{3}\times\mathbb{R}$ can be transformed
into an exotic $S^{3}\times\mathbb{R}$ by end-sum with an exotic
$\mathbb{R}^{4}$. Here we will concentrate on the first construction,
i.e. the exotic $S^{3}\times\mathbb{R}$ is an end of an exotic $\mathbb{R}^{4}$.

Furthermore we will restrict on a subclass of exotic $\mathbb{R}^{4}$
called small exotic $\mathbb{R}^{4}$ (exotic $\mathbb{R}^{4}$ which
can be embedded in a 4-sphere $S^{4}$). For this class there is an
explicit handle decomposition. Small exotic $\mathbb{R}^{4}$'s are
the result of an anomalous behavior in 4-dimensional topology. In
4-manifold topology \cite{Fre:82}, a homotopy-equivalence between
two compact, closed, simply-connected 4-manifolds implies a homeomorphism
between them (a so-called h cobordism). But Donaldson \cite{Don:87}
provided the first smooth counterexample that this homeomorphism is
not a diffeomorphism, i.e. both manifolds are generally not diffeomorphic
to each other. The failure can be localized at some contractible submanifold
(Akbulut cork) so that an open neighborhood of this submanifold is
a small exotic $\mathbb{R}^{4}$. The whole procedure implies that
this exotic $\mathbb{R}^{4}$ can be embedded in the 4-sphere $S^{4}$.
The idea of the construction is simply given by the fact that every
smooth h-cobordism between non-diffeomorphic 4-manifolds can be written
as a product cobordism except for a compact contractible sub-h-cobordism
$V$, the Akbulut cork. An open subset $U\subset V$ homeomorphic
to $[0,1]\times{{\mathbb{R}}^{4}}$ is the corresponding sub-h-cobordism
between two exotic ${{\mathbb{R}}^{4}}$'s. These exotic ${{\mathbb{R}}^{4}}$'s
are called ribbon ${{\mathbb{R}}^{4}}$'s. They have the important
property of being diffeomorphic to open subsets of the standard ${{\mathbb{R}}^{4}}$.
In \cite{DeMichFreedman1992} Freedman and DeMichelis constructed
also a continuous family of small exotic $\mathbb{R}^{4}$. Now we
are ready to discuss the decomposition of a small exotic $\mathbb{R}^{4}$
by Bizaca and Gompf \cite{BizGom:96} using special pieces, the handles
forming a handle body. Every 4-manifold can be decomposed (seen as
handle body) using standard pieces such as $D^{k}\times D^{4-k}$,
the so-called $k$-handle attached along $\partial D^{k}\times D^{4-k}$
to the boundary $S^{3}=\partial D^{4}$ of a $0-$handle $D^{0}\times D^{4}=D^{4}$.
The construction of the handle body for the small exotic $\mathbb{R}^{4}$,
called $R^{4}$ in the following, can be divided into two parts:
\[
R^{4}=A_{cork}\cup_{D^{2}}CH\qquad\mbox{decomposition of small exotic \ensuremath{\mathbb{R}^{4}}}.
\]
The first part is known as the Akbulut cork, a contractable 4-manifold
with boundary a homology 3-sphere (a 3-manifold with the same homology
as the 3-sphere). The Akbulut cork $A_{cork}$ is given by a linking
between a 1-handle and a 2-handle of framing $0$. The second part
is the Casson handle $CH$ which will be considered now. 

Let us start with the basic construction of the Casson handle $CH$.
Let $M$ be a smooth, compact, simple-connected 4-manifold and $f:D^{2}\to M$
a (codimension-2) mapping. By using diffeomorphisms of $D^{2}$ and
$M$, one can deform the mapping $f$ to get an immersion (i.e. injective
differential) generically with only double points (i.e. $\#|f^{-1}(f(x))|=2$)
as singularities \cite{GolGui:73}. But to incorporate the generic
location of the disk, one is rather interesting in the mapping of
a 2-handle $D^{2}\times D^{2}$ induced by $f\times id:D^{2}\times D^{2}\to M$
from $f$. Then every double point (or self-intersection) of $f(D^{2})$
leads to self-plumbings of the 2-handle $D^{2}\times D^{2}$. A self-plumbing
is an identification of $D_{0}^{2}\times D^{2}$ with $D_{1}^{2}\times D^{2}$
where $D_{0}^{2},D_{1}^{2}\subset D^{2}$ are disjoint sub-disks of
the first factor disk. In complex coordinates the plumbing may be
written as $(z,w)\mapsto(w,z)$ or $(z,w)\mapsto(\bar{w},\bar{z})$
creating either a positive or negative (respectively) double point
on the disk $D^{2}\times0$ (the core). Consider the pair $(D^{2}\times D^{2},\partial D^{2}\times D^{2})$
and produce finitely many self-plumbings away from the attaching region
$\partial D^{2}\times D^{2}$ to get a kinky handle $(k,\partial^{-}k)$
where $\partial^{-}k$ denotes the attaching region of the kinky handle.
A kinky handle $(k,\partial^{-}k)$ is a one-stage tower $(T_{1},\partial^{-}T_{1})$
and an $(n+1)$-stage tower $(T_{n+1},\partial^{-}T_{n+1})$ is an
$n$-stage tower union kinky handles $\bigcup_{\ell=1}^{n}(T_{\ell},\partial^{-}T_{\ell})$
where two towers are attached along $\partial^{-}T_{\ell}$. Let $T_{n}^{-}$
be $(\mbox{interior}T_{n})\cup\partial^{-}T_{n}$ and the Casson handle
\[
CH=\bigcup_{\ell=0}T_{\ell}^{-}
\]
is the union of towers (with direct limit topology induced from the
inclusions $T_{n}\hookrightarrow T_{n+1}$).

The main idea of the construction above is very simple: an immersed
disk (disk with self-intersections) can be deformed into an embedded
disk (disk without self-intersections) by sliding one part of the
disk along another (embedded) disk to kill the self-intersections.
Unfortunately the other disk can be immersed only. But the immersion
can be deformed to an embedding by a disk again etc. In the limit
of this process one ''shifts the self-intersections into infinity''
and obtains the standard open 2-handle $(D^{2}\times\mathbb{R}^{2},\partial D^{2}\times\mathbb{R}^{2})$.
In the proof of Freedman \cite{Fre:82}, the main complications come
from the lack of control about this process.

A Casson handle is specified up to (orientation preserving) diffeomorphism
(of pairs) by a labeled finitely-branching tree with base-point {*},
having all edge paths infinitely extendable away from {*}. Each edge
should be given a label $+$ or $-$. Here is the construction: tree
$\to CH$. Each vertex corresponds to a kinky handle; the self-plumbing
number of that kinky handle equals the number of branches leaving
the vertex. The sign on each branch corresponds to the sign of the
associated self plumbing. The whole process generates a tree with
infinitely many levels. In principle, every tree with a finite number
of branches per level realizes a corresponding Casson handle. Each
building block of a Casson handle, the ``kinky'' handle with $n$
kinks, is diffeomorphic to the $n-$times boundary-connected sum $\natural_{n}(S^{1}\times D^{3})$
(see appendix \ref{sec:Connected-and-boundary-connected}) with two
attaching regions. The number of end-connected sums is exactly the
number of self intersections of the immersed two handle. One region
is a tubular neighborhood of band sums of Whitehead links connected
with the previous block. The other region is a disjoint union of the
standard open subsets $S^{1}\times D^{2}$ in $\#_{n}S^{1}\times S^{2}=\partial(\natural_{n}S^{1}\times D^{3})$
(this is connected with the next block).

For the construction of an exotic $S^{3}\times\mathbb{R}$, denoted
by $S^{3}\times_{\theta}\mathbb{R}$, we consider the complement $R^{4}\setminus D^{4}$
or the decomposition
\[
S^{3}\times_{\theta}\mathbb{R}=R^{4}\setminus D^{4}=\left(A_{cork}\setminus D^{4}\right)\cup_{D^{2}}CH
\]
The first part $A_{cork}\setminus D^{4}$ contains a cobordism between
the 3-sphere $S^{3}$ and the boundary of the Akbulut cork $\partial A_{cork}$
(a homology 3-sphere). The complement $\mathbb{R}^{4}\setminus D^{4}$
is conformally equivalent to $S^{3}\times\mathbb{R}$. Equivalently,
the complement $R^{4}\setminus D^{4}$ is diffeomorphic to an exotic
$S^{3}\times\mathbb{R}$. But the exoticness is not confined to a
compact subset but concentrated at infinity (for instance at $+\infty$).
In our case we choose a decomposition like
\[
S^{3}\times_{\theta}\mathbb{R}=M(S^{3},\partial A_{cork})\cup_{D^{2}}CH
\]
where $M(S^{3},\partial A_{cork})$ is a cobordism between $S^{3}$
and $\partial A_{cork}$. For the Casson handle we need another representation
obtained by using Morse theory (see \cite{Mil:63}). Every kinky handle
$(k,\partial^{-}k)$ is given by $n$ pairs of $1-/2-$handle pairs,
where $n$ is the number of kinks (or self-intersections). These handles
are given by the level sets of the Morse functions
\begin{eqnarray*}
f_{1}=x^{2}+y^{2}+z^{2}-t^{2} & \quad\mbox{for\;} & \mbox{the\;\ 1-handle}\\
f_{2}=x^{2}+y^{2}-z^{2}-t^{2} & \quad\mbox{for}\; & \mbox{the\;\ 2-handle}
\end{eqnarray*}
i.e. by the sets $L(f_{i},C)=\left\{ (x,y,z,t)\,|\: f_{i}(x,y,z,t)=C=const.\right\} $
for $i=1,2$. Now we represent the Casson handle by the union of all
$n-$stage towers
\[
CH=\bigcup_{\mbox{level \ensuremath{\ell}\ of tree }\mathcal{T}}(int(T_{\ell})\cup\partial^{-}T_{\ell})
\]
arranged along the tree $\mathcal{T}$. But every tower $T_{\ell}$
is given by the union of pairs $(f_{1},f_{2})$. But what is the geometry
of $T_{\ell}$ (and better of $int(T_{\ell})$)? Every level set $L(f_{1},C)$
and $L(f_{2},C)$ is a hyperbolic 3-manifold (i.e. with negative curvature)
and the union of all level sets is a hyperbolic 4-manifold. A central
point in our argumentation is Mostow rigidity, a central property
of all hyperbolic 3-manifolds (or higher) with finite volume explained
in the next subsection.

\subsection{The hyperbolic geometry of $CH$\label{sub:hyperbolic-geometry-CH}}

The central element in the Casson handle is a pair of 1- and 2-handles
representing a kinky handle. As we argued above this pair admits a
hyperbolic geometry (or it is a hyperbolic 3-manifold) having negative
scalar curvature. A 3-manifold admits a hyperbolic structure in the
interior if there is a diffeomorphism to $\mathbb{H}^{3}/\Gamma$
where $\Gamma$ is a discrete subgroup $\Gamma\subset SO(3,1)$ of
the Lorentz group and we have a representation of the fundamental
group $\pi_{1}(M)$ into $SO(3,1)$ (the isometry group of the hyperbolic
space $\mathbb{H}^{3}$. One property of hyperbolic 3- and 4-manifolds
is central: \noun{Mostow rigidity}. As shown by Mostow \cite{Mos:68},
every hyperbolic $n-$manifold $n>2$ with finite volume has this
property: \emph{Every diffeomorphism (especially every conformal transformation)
of a hyperbolic $n-$manifold with finite volume is induced by an
isometry.} Therefore one cannot scale a hyperbolic 3-manifold with
finite volume. Then the volume $vol(\:)$ and the curvature are topological
invariants but for later usages we combine the curvature and the volume
into the Chern-Simons invariant $CS(\:)$. But more is true: in a
hyperbolic 3-manifold there are special surfaces which cannot be contracted,
called incompressible surface. A properly embedded connected surface
$S\subset N$ in a 3-manifold $N$ is called 2-sided if its normal
bundle is trivial, and 1-sided if its normal bundle is nontrivial.
The \textquoteleft{}sides\textquoteright{} of $S$ then correspond
to the components of the complement of $S$ in a tubular neighborhood
$S\times[0,1]\subset N$. A 2-sided connected surface $S$ other than
$S^{2}$ or $D^{2}$ is called incompressible if for each disk $D\subset N$
with $D\cap S=\partial D$ there is a disk $D'\subset S$ with $\partial D\lyxmathsym{\textasciiacute}=\partial D$,
i.e. the boundary of the disk $D$ can be contracted in the surface
$S$. The boundary of a 3-manifold is an incompressible surface. More
importantly, this surface can be detected in the fundamental group
$\pi_{1}(N)$ of the 3-manifold, i.e. there is an injective homomorphism
$\pi_{1}(S)\to\pi_{1}(N)$. The consequence of all properties is the
following conclusion:\\
\emph{The tower $T_{\ell}$ has a hyperbolic geometry (with finite
volume) and therefore fixed size, i.e. it cannot be scaled by any
diffeomorphism or conformal transformation. Then we obtain an invariant
decomposition of the Casson handle into towers arranged with respect
to a tree. Secondly, inside of every tower $T_{\ell}$ there is (at
least one) an incompressible surface also of fixed size.}\\
In case of the tower $T_{\ell}$, one knows two incompressible surfaces,
the two tori coming from the complement of the Whitehead link (with
two components) used in the construction.

\subsection{The path integral of the exotic $S^{3}\times\mathbb{R}$}

Now we will discuss the path integral using the decomposition
\[
S^{3}\times_{\theta}\mathbb{R}=M(S^{3},\partial A_{cork})\cup_{D^{2}}\left(\bigcup_{\mbox{level \ensuremath{\ell}\ of tree }\mathcal{T}}(int(T_{\ell})\cup\partial^{-}T_{\ell})\right)
\]
and we remark that the construction of the cobordism $M(S^{3},\partial A_{cork})$
requires the usage of a Casson handle again, denoted by $M(S^{3},\partial A_{cork})\cup CH_{cork}$.
Therefore we have to clarify the role of the Casson handle. In the
previous subsection \ref{sub:hyperbolic-geometry-CH}, we discussed
the strong connection between geometry and topology for hyperbolic
manifolds. The topology of $S^{3}\times_{\theta}\mathbb{R}$ is rather
trivial but the smoothness structure (and therefore the differential
topology) can be very complicate. 

As stated above, the boundary terms can be factorized from the terms
in the interior. Formally we obtain
\[
Z(S^{3}\times_{\theta}\mathbb{R})=\left\{ \prod_{\ell}Z(\partial^{-}T_{\ell})Z(\partial A_{cork})Z(S^{3})\right\} \left(\prod_{\ell}Z(int(T_{\ell})\right)Z(CH_{cork})
\]
and for an expectation value of the observable $\mathcal{O}$
\[
\langle S^{3}\times_{\theta}\mathbb{R}|\,\mathcal{O}\,|S^{3}\times_{\theta}\mathbb{R}\rangle
\]
but for the following we have to discuss it more fully. To understand
the time-like evolution of a disk (or a surface), we have to describe
a disk inside of a Casson handle as pioneered by Bizaca \cite{Bizaca1995}.
With the same arguments, one can also describe the modification of
the 3-sphere into homology 3-spheres $\Sigma$. But then we obtain
(formally) an infinite sequence of homology 3-spheres $\Sigma_{1}\to\Sigma_{2}\to\cdots$
with amplitudes
\[
Z(S^{3}\times_{\theta}\mathbb{R})=\langle\Sigma_{1}|T_{M}|\Sigma_{2}\rangle\langle\Sigma_{2}|T_{M}|\Sigma_{3}\rangle\cdots
\]
including the boundary terms. Every spatial section $\Sigma_{n}$
can be seen as an element of the phase space in quantum gravity. Therefore
this change of transitions is a \textbf{topological phase transition}
which will be further investigated in our work.

The choice of the boundary term has a kind of arbitrariness. We can
choose the decomposition much finer to get more boundary terms. Therefore
the path integral (\ref{eq:path-integral-boundary}) must be extended
away from the boundary. We will discuss this extension also in our
forthcoming work.

Before we go ahead we have to discuss the foliation structure of $S^{3}\times_{\theta}\mathbb{R}$
or the appearance of different time variables. As stated above, our
space-time has the topology of $S^{3}\times\mathbb{R}$ with equal
slices parametrized by a \textbf{topological time $t_{TOP}$}, i.e.
\[
\left(S^{3}\times_{\theta}\mathbb{R}\right)_{TOP}=\left\{ (p,t_{TOP})\,|\, p\in S^{3},t_{TOP}\in\mathbb{R}\right\} =\left\{ S^{3}\times\left\{ t_{TOP}\right\} |t_{TOP}\in\left\{ -\infty\ldots+\infty\right\} \right\} 
\]
defined by the topological embedding $S^{3}\hookrightarrow S^{3}\times_{\theta}\mathbb{R}$.
It is the defining property of exotic smoothness that $S^{3}$ inside
of $S^{3}\times_{\theta}\mathbb{R}$ is only a topological 3-sphere,
i.e. it is wildly embedded and so only represented by an infinite
polyhedron. There is another possibility to introduce $t_{TOP}$ which
will point us to the smooth case. For that purpose we define a map
$F:S^{3}\times\mathbb{R}\to\mathbb{R}$ by $(x,t)\mapsto t$ so that
$t_{TOP}=F(p)$ for $p\in S^{3}\times\mathbb{R}$. In contrast one
also has the smooth time $t_{Diff}$ which we have to define now.
Locally it is the smooth (physical coordinate) time. We know also
that the exotic $S^{3}\times_{\theta}\mathbb{R}$ is composed by a
sequence $\Sigma_{1}\to\Sigma_{2}\to\Sigma_{3}\to\cdots$ of homology
3-spheres or better by a sequence $M(\Sigma_{1},\Sigma_{2})\cup_{\Sigma_{2}}M(\Sigma_{2},\Sigma_{3})\cup_{\Sigma_{3}}\ldots$
of (homology) cobordism between the homology 3-spheres. All sequences
are ordered and so it is enough to analyze one cobordism $M(\Sigma_{1},\Sigma_{2})$.
Every cobordism between two homology 3-spheres $\Sigma_{1}$ and $\Sigma_{2}$
is characterized by the existence of a finite number of 1-/2-handle
pairs (or dually 2-/3-handle pairs). Now we define a smooth map $F_{cob}:M(\Sigma_{1},\Sigma_{2})\to[0,1]$
which must be a Morse function (i.e. it has isolated critical points)
\cite{Mil:63}. The number of critical points $N$ of $F$ is even,
say $N=2k$ where $k$ is the number of 1-/2-handle pairs. These critical
points are also denoted as naked singularities in GR (but of bounded
curvature). Like in the case of topological time $t_{TOP}$ we introduce
the \textbf{smooth time} by $t_{Diff}=F_{cob}(p)$ for all $p\in M(\Sigma_{1},\Sigma_{2})\subset S^{3}\times_{\theta}\mathbb{R}$.
The extension of $t_{Diff}$ to the whole $S^{3}\times_{\theta}\mathbb{R}$
by the Morse function $F:S^{3}\times_{\theta}\mathbb{R}\to\mathbb{R}$
is straightforward $t_{Diff}=F(p)$ for all $p\in S^{3}\times_{\theta}\mathbb{R}$.
The cobordism $M(\Sigma_{1},\Sigma_{2})$ is part of the exotic $S^{3}\times_{\theta}\mathbb{R}$
and can be embedded to make it $S^{3}\times[0,1]$ topologically.
Therefore function $F_{cob}$ is a continuous function which is strictly
increasing on future directed causal curves, so it is a time function
(see \cite{BernalSanchez2006,BernalSanchez2007}). But there is also
another method to construct $t_{Diff}$ by using codimension-1 foliations.
In \cite{AsselmeyerKrol2009} we uncovered a strong relation between
codimension-1 foliations (also used to construct a Lorentz structure
on a manifold) and exotic smoothness structures for a small exotic
$\mathbb{R}^{4}$. The coordinate of this codimension-1 submanifold
is also the smooth time $t_{Diff}$. This approach will be more fully
discussed in our forthcoming work.

\section{The expectation value of the area and inflation}

In \ref{sub:hyperbolic-geometry-CH} we described the hyperbolic geometry
originated in the exotic smoothness structure of $S^{3}\times_{\theta}\mathbb{R}$.
Because of this hyperbolic geometry, there are incompressible surfaces
inside of the hyperbolic manifold as the smallest possible units of
geometry. Then Mostow rigidity determines the behavior of this incompressible
surface. At first we will concentrate on the first cobordism $M(S^{3},\partial A_{cork})$
between $S^{3}$ and the boundary $\partial A_{cork}$ of the Akbulut
cork. The area of a surface is given by
\[
A(e,S)=\intop_{S}d^{2}\sigma\sqrt{E^{a}E^{b}n_{a}n_{b}}
\]
with the normal vector $n_{a}$ and the densitized frame $E^{a}=\det(e)\, e^{a}$.
The expectation value of the area $A$ 
\[
\langle S^{3}|\, A(e,S)\,|\partial A_{cork}\rangle=\frac{1}{Z(M(S^{3},\partial A_{cork}))}\intop De\: A(e,S)\,\exp\left(\frac{i}{\hbar}S_{EH}[e,M(S^{3},\partial A_{cork}]\right)
\]
depends essentially on the hyperbolic geometry. As argued above, this
cobordism has a hyperbolic geometry but in the simplest case, the
boundary of the Akbulut cork is the homology 3-sphere $\partial A_{cork}=\Sigma(2,5,7)$,
a Brieskorn homology 3-sphere. Now we study the area of a surface
where one direction is along the time axis. Then we obtain a decomposition
of the surface into a sum of small surfaces so that every small surface
lies in one component of the cobordism. Remember, that the cobordism
$M(S^{3},\partial A_{cork})$ is decomposed into the trivial cobordism
$S^{3}\times[0,1]$ and a Casson handle $CH=\cup_{\ell}T_{\ell}$.
Then the decomposition of the surface
\[
S=\cup_{\ell}S_{\ell}
\]
corresponds to the decomposition of the expectation value of the area
\[
A_{\ell}(e,S_{\ell})=\intop_{S_{\ell}}d^{2}\sigma\sqrt{E^{a}E^{b}n_{a}n_{b}}
\]
so that
\[
\langle\partial^{-}T_{\ell}|A_{\ell'}(e,S_{\ell'})|\partial^{-}T_{\ell+1}\rangle=\langle\partial^{-}T_{\ell}|A_{\ell}(e,S_{\ell})|\partial^{-}T_{\ell+1}\rangle\delta_{\ell\ell'}
\]
and
\[
\langle S^{3}|\, A(e,S)\,|\partial A_{cork}\rangle=\sum_{\ell}\langle\partial^{-}T_{\ell}|A_{\ell}(e,S_{\ell})|\partial^{-}T_{\ell+1}\rangle
\]
The initial value for $\ell=0$ is the expectation value
\[
\langle\partial^{-}T_{0}|A_{0}(e,S_{0})|\partial^{-}T_{1}\rangle=a_{0}^{2}
\]
where $a_{0}$is the radius of the 3-sphere $S^{3}$. But because
of the hyperbolic geometry (with constant curvature because of Mostow
rigidity) every further level scales this expectation value by a constant
factor. Therefore, to calculate the expectation value, we have to
study the scaling behavior.

Consider a cobordism $M(\Sigma_{0},\Sigma_{1})$ between the homology
3-spheres $\Sigma_{0},\Sigma_{1}$. As shown by Witten \cite{Wit:89.2,Wit:89.3,Wit:91.2},
the action
\begin{equation}
\intop_{\Sigma_{0,1}}\,^{3}R\sqrt{h}\, d^{3}x=L\cdot CS(\Sigma_{0,1})\label{eq:Witten-relation}
\end{equation}
for every 3-manifold (in particular for $\Sigma_{0}$ and $\Sigma_{1}$
denoted by $\Sigma_{0,1}$) is related to the Chern-Simons action
$CS(\Sigma_{0,1})$ (defined in \ref{sec:Chern-Simons-invariant}).
The scaling factor $L$ is related to the volume by $L=\sqrt[3]{vol(\Sigma_{0,1})}$
and we obtain formally
\begin{equation}
L\cdot CS(\Sigma_{0,1},A)=L^{3}\cdot\frac{CS(\Sigma_{0,1})}{L^{2}}=\intop_{\Sigma_{0,1}}\frac{CS(\Sigma_{0,1})}{L^{2}}\sqrt{h}\, d^{3}x\label{eq:CS-integral-relation}
\end{equation}
by using
\[
L^{3}=vol(\Sigma_{0,1})=\intop_{\Sigma_{0,1}}\sqrt{h}\, d^{3}x\,.
\]
Together with 
\[
^{3}R=\frac{3k}{a^{2}}
\]
one can compare the kernels of the integrals of (\ref{eq:Witten-relation})
and (\ref{eq:CS-integral-relation}) to get for a fixed time 
\[
\frac{3k}{a^{2}}=\frac{CS(\Sigma_{0,1})}{L^{2}}\,.
\]
This gives the scaling factor
\begin{equation}
\vartheta=\frac{a^{2}}{L^{2}}=\frac{3}{CS(\Sigma_{0,1})}\label{eq:scaling-CH}
\end{equation}
where we set $k=1$ in the following. The hyperbolic geometry of the
cobordism is best expressed by the metric

\begin{equation}
ds^{2}=dt^{2}-a(t)^{2}h_{ik}dx^{i}dx^{k}\label{eq:FRW-metric}
\end{equation}
also called the Friedmann-Robertson-Walker metric (FRW metric) with
the scaling function $a(t)$ for the (spatial) 3-manifold. But Mostow
rigidity enforces us to choose 
\[
\left(\frac{\dot{a}}{a}\right)^{2}=\frac{1}{L^{2}}
\]
in the length scale $L$ of the hyperbolic structure. But why is it
possible to choose the FRW metric? At first we state that the FRW
metric is not sensitive to the topology of the space-time. One needs
only a space-time which admits a slicing with respect to a smooth
time $t_{Diff}$ and a metric of constant curvature for every spatial
slice. Then for the cobordims $M(\Sigma_{1},\Sigma_{2})$ between
$\Sigma_{1}$ and $\Sigma_{2}$we have two cases: the curvature parameter
$k(\Sigma_{1})$ of $\Sigma_{1}$ (say $k(\Sigma_{1})=+1$) jumps
to the value $k(\Sigma_{2})$ of $\Sigma_{2}$ (say $k(\Sigma_{2})=-1$)
or both curvatures remain constant. The second case is the usual one.
Each homology 3-sphere $\Sigma_{1},\Sigma_{2}$ has the same geometry
(or geometric structure in the sense of Thurston \cite{Thu:97}) which
is hyperbolic in most case. The first case is more complicated. Here
we need the smooth function to represent the jump in the curvature
parameter $k$. Lets choose the function $k:\mathbb{R}\to\mathbb{R}$
\[
t\mapsto\left\lbrace \begin{array}{cc}
+1 & 0\leq t\\
1-2\cdot\exp\left(-\lambda\cdot t^{-2}\right) & t>0
\end{array}\right.
\]
which is smooth and the parameter $\lambda$ determines the slope
of this function. Furthermore the metric (\ref{eq:FRW-metric}) is
also the metric of a hyperbolic space (which has to fulfill Mostow
rigidity because the cobordism $M(\Sigma_{1},\Sigma_{2})$ is compact). 

In the following we will switch to quadratic expressions because we
will determine the expectation value of the area. Then we obtain 
\begin{equation}
da^{2}=\frac{a^{2}}{L^{2}}\, dt^{2}=\vartheta\, dt^{2}\label{eq:scale-quadratic-expansion}
\end{equation}
with respect to the scale $\vartheta$. By using the tree of the Casson
handle, we obtain a countable infinite sum of contributions for (\ref{eq:scale-quadratic-expansion}).
Before we start we will clarify the geometry of the Casson handle.
The discussion of the Morse functions above uncovers the hyperbolic
geometry of the Casson handle (see also the subsection \ref{sub:hyperbolic-geometry-CH}).
Therefore the tree corresponding to the Casson handle must be interpreted
as a metric tree with hyperbolic structure in $\mathbb{H}^{2}$ and
metric $ds^{2}=(dx^{2}+dy^{2})/y^{2}$. The embedding of the Casson
handle in the cobordism is given by the rules
\begin{enumerate}
\item The direction of the increasing levels $n\to n+1$ is identified with
$dy^{2}$ and $dx^{2}$ is the number of edges for a fixed level with
scaling parameter $\vartheta$. 
\item The contribution of every level in the tree is determined by the previous
level best expressed in the scaling parameter $\vartheta$. 
\item An immersed disk at level $n$ needs at least one disk to resolve
the self-intersection point. This disk forms the level $n+1$ but
this disk is connected to the previous disk. So we obtain for $da^{2}|_{n+1}$
at level $n+1$
\[
da^{2}|_{n+1}\sim\vartheta\cdot da^{2}|_{n}
\]
up to a constant. 
\end{enumerate}
By using the metric $ds^{2}=(dx^{2}+dy^{2})/y^{2}$ with the embedding
($y^{2}\to n+1$, $dx^{2}\to\vartheta$) we obtain for the change
$dx^{2}/y^{2}$ along the $x-$direction (i.e. for a fixed $y$) $\frac{\vartheta}{n+1}$.
This change determines the scaling from the level $n$ to $n+1$,
i.e. 
\[
da^{2}|_{n+1}=\frac{\vartheta}{n+1}\cdot da^{2}|_{n}=\frac{\vartheta^{n+1}}{(n+1)!}\cdot da^{2}|_{0}
\]
and after the whole summation (as substitute for an integral for the
discrete values) we obtain for the relative scaling 
\begin{equation}
a^{2}=\sum_{n=0}^{\infty}\left(da^{2}|_{n}\right)=a_{0}^{2}\cdot\sum_{n=0}^{\infty}\frac{1}{n!}\vartheta^{n}=a_{0}^{2}\cdot\exp\left(\vartheta\right)=a_{0}^{2}\cdot l_{scale}\label{eq:scaling}
\end{equation}
with $da^{2}|_{0}=a_{0}^{2}$. With this result in mind, we consider
the expectation value where we use the constant scalar curvature (Mostow
rigidity). By using the normalization, many terms are neglected (like
the boundary terms). 
\begin{eqnarray*}
\langle S^{3}|\, A(e,S)\,|\partial A_{cork}\rangle & = & \frac{\left\{ \prod_{\ell}Z(\partial^{-}T_{\ell})Z(\partial A_{cork})Z(S^{3})\right\} \sum_{n=0}^{\infty}\langle\partial^{-}T_{n}|A_{n}(e,S_{n})|\partial^{-}T_{n+1}\rangle}{\left\{ \prod_{\ell}Z(\partial^{-}T_{\ell})Z(\partial A_{cork})Z(S^{3})\right\} }\\
 & = & \sum_{n=0}^{\infty}\langle\partial^{-}T_{n}|A_{n}(e,S_{n})|\partial^{-}T_{n+1}\rangle
\end{eqnarray*}
Finally we obtain for the area $a_{0}^{2}$ for the first level $\ell=0$
\begin{eqnarray*}
\langle S^{3}|\, A(e,S)\,|\partial A_{cork}\rangle & = & \sum_{n=0}^{\infty}\langle\partial^{-}T_{n}|A_{n}(e,S_{n})|\partial^{-}T_{n+1}\rangle\\
 & = & a_{0}^{2}\cdot\sum_{n=0}^{\infty}\frac{1}{n!}\left(\frac{3}{CS(\partial A_{cork})}\right)^{n}\\
 & = & a_{0}^{2}\cdot\exp\left(\frac{3}{CS(\partial A_{cork})}\right)\,.
\end{eqnarray*}
with the radius $a_{0}$ of $\Sigma_{0}$ and arrive at $a$ for $\Sigma_{1}$.
From the physical point of view we obtain an exponential increase
of the area, i.e. we get an inflationary behavior. This derivation
can be also extended to the next Casson handle but we have to determine
the 3-manifold in which $\partial A_{cork}$ can change. It will be
done below.

\section{An effective theory}

Now will ask for an effective theory where the influence of the exotic
smoothness structure is contained in some moduli (or some field).
As explained above the main characteristics is given by a change of
the (spatial) 4-manifold (but without changing the homology). Therefore
let us describe this change (a so-called homology cobordism) between
two homology 3-spheres $\Sigma_{0}$ and $\Sigma_{1}$. The situation
can be described by a diagram 
\begin{eqnarray}
\Sigma_{1} & \stackrel{\Psi}{\longrightarrow} & \mathbb{R}\nonumber \\
\phi\downarrow & \circlearrowright & \updownarrow id\label{eq:commuting-diagram}\\
\Sigma_{0} & \stackrel{\psi}{\longrightarrow} & \mathbb{R}\nonumber 
\end{eqnarray}
which commutes. The two functions $\psi$ and $\Psi$ are the Morse
function of $\Sigma_{0}$ and $\Sigma_{1}$, respectively, with $\Psi=\psi\circ\phi$.
The Morse function over $\Sigma_{0,1}$ is a function $\Sigma_{0,1}\to\mathbb{R}$
having only isolated, non-degenerated, critical points (i.e. with
vanishing first derivatives at these points). A homology 3-sphere
has two critical points (located at the two poles). The Morse function
looks like $\pm||x||^{2}$ at these critical points. The transition
$y=\phi(x)$ represented by the (homology) cobordism $M(\Sigma_{0},\Sigma_{1})$
maps the Morse function $\psi(y)=||y||^{2}$ on $\Sigma_{0}$ to the
Morse function $\Psi(x)=||\phi(x)||^{2}$ on $\Sigma_{1}$. The function
$-||\phi||^{2}$ represents also the critical point of the cobordism
$M(\Sigma_{0},\Sigma_{1})$. But as we learned above, this cobordism
has a hyperbolic geometry and we have to interpret the function $||\phi(x)||^{2}$
not as Euclidean form but change it to the hyperbolic geometry so
that
\[
-||\phi||^{2}=-\left(\phi_{1}^{2}+\phi_{2}^{2}+\phi_{3}^{2}\right)\to-e^{-2\phi_{1}}(1+\phi_{2}^{2}+\phi_{3}^{2})
\]
i.e. we have a preferred direction represented by a single scalar
field $\phi_{1}:\Sigma_{1}\to\mathbb{R}$. Therefore, the transition
$\Sigma_{0}\to\Sigma_{1}$ is represented by a single scalar field
$\phi_{1}:\Sigma_{1}\to\mathbb{R}$ and we identify this field as
the moduli. Finally we interpret this Morse function in the interior
of the cobordism $M(\Sigma_{0},\Sigma_{1})$ as the potential (shifted
away from the point $0$ ) of the scalar field $\phi$ with Lagrangian
\[
L=R+(\partial_{\mu}\phi)^{2}-\frac{\rho}{2}(1-\exp\left(-\lambda\phi\right))^{2}
\]
with two free constants $\rho$ and $\lambda$. For the value $\lambda=\sqrt{2/3}$
and $\rho=3M^{2}$ we obtain the Starobinski model \cite{Starobinski1980}
(by a conformal transformation using $\phi$ and a redefinition of
the scalar field \cite{Whitt1984})
\begin{equation}
L=R+\frac{1}{6M^{2}}R^{2}\label{eq:Starobinski-model}
\end{equation}
with the mass scale $M\ll M_{P}$ much smaller than the Planck mass.
From our discussion above, the appearance of this model is not totally
surprising. It favors a surface to be incompressible (which is compatible
with the properties of hyperbolic manifolds). In the next section
we will determine this mass scale.

\section{A cosmological model compared to the Planck satellite results}

In this section we will go a step further and discuss the path integral
for $S^{3}\times\mathbb{R}$ where we sum over all smoothness structures.
Furthermore we will assume that $S^{3}\times\mathbb{R}$ is the end
of a small exotic $\mathbb{R}^{4}$. But then we have to discuss the
parametrization of all Casson handles. As discussed by Freedman \cite{Fre:82},
all Casson handles can be parametrized by a dual tree where the vertices
are 5-stage towers (with three extra conditions). We refer to \cite{Fre:82}
or to \cite{FreQui:90} for the details of the well-known construction.
This tree has one root from which two 5-stage towers branch. Every
tower has an attaching circle of any framing. Using Bizacas technique
\cite{Bizaca1995}, we obtain an attaching of a 5-tower along the
sum $P\#P$ of two Poincare spheres $P$ (for the two towers). Therefore
for the universal case, we obtain two transitions
\[
S^{3}\stackrel{cork}{\longrightarrow}\partial A_{cork}\stackrel{tower}{\longrightarrow}P\#P
\]
with the scaling behavior
\[
a=a_{0}\cdot\exp\left(\frac{3}{2\cdot CS(\partial A_{cork})}+\frac{3}{2\cdot CS(P\#P)}\right)\,.
\]
It can be expressed by the expectation value
\[
\langle S^{3}|\, A(e,S)\,|P\#P\rangle=a_{0}^{2}\cdot\exp\left(\frac{3}{CS(\partial A_{cork})}+\frac{3}{CS(P\#P)}\right)
\]
for the transition $S^{3}\to P\#P$. It is important to note, that
this expectation value is the sum over all smoothness structures of
$S^{3}\times\mathbb{R}$ and we obtain also
\begin{eqnarray*}
\langle S^{3}\times\mathbb{R}|\: A(e,S)\,|S^{3}\times\mathbb{R}\rangle & = & \sum_{diff\, structures}\langle S^{3}\times_{\theta}\mathbb{R}|\: A(e,S)\,|S^{3}\times_{\theta}\mathbb{R}\rangle\\
 & = & a_{0}^{2}\cdot\exp\left(\frac{3}{CS(\partial A_{cork})}+\frac{3}{CS(P\#P)}\right)
\end{eqnarray*}
with $a_{0}^{2}$ as the size of the 3-sphere $S^{3}$ at $-\infty$.
With the argumentation above, the smoothness structure has a kind
of universality so that the two transitions above are generic.

In our model (using the exotic smoothness structure), we obtain two
inflationary phases. In the first phase we have a transition
\[
S^{3}\to\partial A_{cork}
\]
and for the simplest case $\partial A{}_{cork}=\Sigma(2,5,7)$, a
Brieskorn homology 3-sphere. Now we will assume that the 3-sphere
has Planck-size
\[
a_{0}=L_{P}=\sqrt{\frac{hG}{c^{3}}}
\]
then we obtain for the size
\[
a=L_{P}\cdot\exp\left(\frac{3}{2\cdot CS(\Sigma(2,5,7)}\right)\,.
\]
We can use the method of Fintushel and Stern \cite{FinSte:90,KirKla:90,FreGom:91}
to calculate the Chern-Simons invariants for the Brieskorn spheres.
The calculation can be found in \ref{sec:Chern-Simons-invariant-of-B}.
Note, that the relation (\ref{eq:Witten-relation}) is only true for
the Levi-Civita connection. Then the Chern-Simons invariant is uniquely
defined to be the minimum, denoted by $\tau()$ (see \ref{CS-invariante}).
then we obtain for the invariant (\ref{eq:CS-for-B}) so that 
\[
L_{P}\cdot\exp\left(\frac{140}{3}\right)\approx7.5\cdot10^{-15}m
\]
is the size of the cosmos at the end of the first inflationary phase.
This size can be related to an energy scale by using it as Compton
length and one obtains 165 MeV, comparable to the energy scale of
the QCD. For the two inflationary transitions
\[
S^{3}\to\Sigma(2,5,7)\to P\#P
\]
one obtains the size
\[
a=L_{P}\cdot\exp\left(\frac{140}{3}+90\right)\approx9.14\cdot10^{24}m\approx10^{9}Lj\:.
\]
As explained above, the effective theory is the Starobinsky model.
This model is in very good agreement with results of the Planck satellite
\cite{PlanckInflation2013} with the two main observables
\begin{eqnarray*}
n_{s}\sim & 0.96 & \mbox{ spectral index for scalar perturbations}\\
r\sim & 0.004 & \mbox{ tensor-to-scalar ratio}
\end{eqnarray*}
but one parameter of the model is open, the energy scale $M$ in Planck
units. In our model it is related to the second derivative of the
Morse function, which is the curvature of the critical point. In our
paper \cite{AsselmKrol-2013b}, we determined also the energy scale
of the inflation by using a simple argument to incorporate only the
first 3 levels of the Casson handle. For the scale 
\[
\vartheta=\frac{3}{2\cdot CS(\Sigma(2,5,7))}
\]
of the first transition, we obtain the scaling of the Planck energy
(associated to the Planck-sized 3-sphere at the beginning)
\[
E_{Inflation}=\frac{E_{Planck}}{\left(1+\vartheta+\frac{\vartheta^{2}}{2}+\frac{\vartheta^{3}}{6}\right)}
\]
with the relative scaling
\[
\alpha=\frac{E_{Inflation}}{E_{Planck}}=\frac{1}{\left(1+\vartheta+\frac{\vartheta^{2}}{2}+\frac{\vartheta^{3}}{6}\right)}\thickapprox5.5325\cdot10^{-5}
\]
leading to the energy scale of the inflation
\[
E_{Inflation}\thickapprox6.7547\cdot10^{14}GeV
\]
by using $E_{Planck}\approx1.2209\cdot10^{19}GeV$. We remark that
the relative scaling $\alpha\thickapprox5.5325\cdot10^{-5}$ above
is the factor $\alpha=1/6M^{2}$ in the Starobinski model (in agreement
with measurements). Now we can go a step further and discuss the appearance
of the cosmological constant.

Again we can use the hyperbolic geometry to state that the curvature
is negative and we have the Mostow rigidity, i.e. the scalar curvature
of the 4-manifold has a constant value, the cosmological constant
$\Lambda$. If we assume that the 3-sphere has the size of the Planck
length (as above) then we obtain 
\[
\Lambda=\frac{1}{L_{P}^{2}}\cdot\exp\left(-\frac{3}{CS(\Sigma(2,5,7))}-\frac{3}{CS(P\#P)}\right)\,.
\]
With the values of the Chern-Simons invariants (\ref{eq:CS-for-B}),
we obtain the value
\[
\Lambda\cdot L_{P}^{2}=\exp\left(-\frac{280}{3}-180\right)\approx5\cdot10^{-118}
\]
in Planck units. In cosmology, one usually relate the cosmological
constant to the Hubble constant $H_{0}$ (expressing the critical
density) leading to the length scale
\[
L_{c}^{2}=\frac{c^{2}}{3H_{0}^{2}}\,.
\]
The corresponding variable is denoted by $\Omega_{\Lambda}$ and we
obtain
\begin{equation}
\Omega_{\Lambda}=\frac{c^{5}}{3hGH_{0}^{2}}\cdot\exp\left(-\frac{3}{CS(\Sigma(2,5,7))}-\frac{3}{CS(P\#P)}\right)\label{eq:dark-energy}
\end{equation}
in units of the critical density. This formula is in very good agreement
with the WMAP results, i.e. by using the value for the Hubble constant
\[
H_{0}=74\,\frac{km}{s\cdot Mpc}
\]
we are able to calculate the dark energy density to be
\[
\Omega_{\Lambda}=0.729
\]
agreeing with the WMAP results. But it differs from the Planck results
\cite{PlanckCosmoParameters2013} of the Hubble constant 
\[
\left(H_{0}\right)_{Planck}=68\,\frac{km}{s\cdot Mpc}
\]
for which we obtain
\[
\Omega_{\Lambda}\approx0.88
\]
in contrast with the measured value of the dark energy
\[
\left(\Omega_{\Lambda}\right)_{Planck}=0.683\,.
\]
But there is another possibility for the size of the 3-sphere at the
beginning and everything depends on this choice. But we can use the
entropy formula of a Black hole in Loop quantum gravity
\[
S=\frac{A\cdot\gamma_{0}}{4\cdot\gamma\cdot L_{P}^{2}}\cdot2\pi
\]
with
\[
\gamma_{0}=\frac{ln(2)}{\pi\cdot\sqrt{3}}
\]
according to \cite{AshBaezCorichKrasnov1998:blackholeentropie} with
the Immirzi parameter $\gamma$ where the extra factor $2\pi$ is
given by a different definition of $L_{P}$ replacing $h$ by $\hbar$.
In the original approach of Ashtekar in Loop quantum gravity one usually
set $\gamma=1$. If we take it seriously then we obtain a reduction
of the length in (\ref{eq:dark-energy}) 
\[
\frac{1}{L_{P}^{2}}\to\frac{1}{L_{P}^{2}}\cdot\frac{2\cdot ln(2)}{\sqrt{3}}=\frac{2\pi\gamma_{0}}{L_{P}^{2}}\approx0.80037\cdot\frac{1}{L_{P}^{2}}
\]
or the new closed formula
\begin{equation}
\Omega_{\Lambda}=\frac{c^{5}}{3\hbar GH_{0}^{2}}\cdot\gamma_{0}\cdot\exp\left(-\frac{3}{CS(\Sigma(2,5,7))}-\frac{3}{CS(P\#P)}\right)\label{eq:corrected-dark-energy}
\end{equation}
correcting the value $\Omega_{\Lambda}\approx0.88$ to
\[
\Omega_{\Lambda}\approx0.704\quad.
\]
But $\gamma_{0}$ depends on the gauge group and if one uses the value
\cite{Lewandoski2004:entropyblackhole3}
\[
\gamma_{0}=\frac{ln(3)}{\pi\cdot\sqrt{8}}
\]
agreeing also with calculations in the spin foam models \cite{Ansari2008:entropieblackhole2}
then one gets a better fit
\[
\Omega_{\Lambda}\approx0.6836
\]
which is in good agreement with the measurements.

\section{Conclusion}

The strong relation between hyperbolic geometry (of the space-time)
and exotic smoothness is one of the main results in this paper. Then
using Mostow rigidity, geometric observables like area and volume
or curvature are topological invariants which agree with the expectation
values of these observables (calculated via the path integral). We
compared the results with the recent results of the Planck satellite
and found a good agreement. In particular as a direct result of the
hyperbolic geometry, the inflation can be effectively described by
the Starobinski model. Furthermore we also obtained a cosmological
model which produces a realistic cosmological constant.

\section*{Acknowledgement}

The authors acknowledged all critical remarks of the referees increasing
the readability of the paper.

\appendix

\section{Connected and boundary-connected sum of manifolds\label{sec:Connected-and-boundary-connected}}

Now we will define the connected sum $\#$ and the boundary connected
sum $\natural$ of manifolds. Let $M,N$ be two $n$-manifolds with
boundaries $\partial M,\partial N$. The \emph{connected sum} $M\#N$
is the procedure of cutting out a disk $D^{n}$ from the interior
$int(M)\setminus D^{n}$ and $int(N)\setminus D^{n}$ with the boundaries
$S^{n-1}\sqcup\partial M$ and $S^{n-1}\sqcup\partial N$, respectively,
and gluing them together along the common boundary component $S^{n-1}$.
The boundary $\partial(M\#N)=\partial M\sqcup\partial N$ is the disjoint
sum of the boundaries $\partial M,\partial N$. The \emph{boundary
connected sum} $M\natural N$ is the procedure of cutting out a disk
$D^{n-1}$ from the boundary $\partial M\setminus D^{n-1}$ and $\partial N\setminus D^{n-1}$
and gluing them together along $S^{n-2}$ of the boundary. Then the
boundary of this sum $M\natural N$ is the connected sum $\partial(M\natural N)=\partial M\#\partial N$
of the boundaries $\partial M,\partial N$.

\section{Chern-Simons invariant\label{sec:Chern-Simons-invariant}}

Let $P$ be a principal $G$ bundle over the 4-manifold $M$ with
$\partial M\not=0$. Furthermore let $A$ be a connection in $P$
with the curvature 
\[
F_{A}=dA+A\wedge A
\]
and Chern class
\[
C_{2}=\frac{1}{8\pi^{2}}\int\limits _{M}tr(F_{A}\wedge F_{A})
\]
for the classification of the bundle $P$. By using the Stokes theorem
we obtain 
\begin{equation}
\int\limits _{M}tr(F_{A}\wedge F_{A})=\int\limits _{\partial M}tr(A\wedge dA+\frac{2}{3}A\wedge A\wedge A)\label{eq:stokes-CS}
\end{equation}
with the Chern-Simons invariant 
\begin{equation}
CS(\partial M,A)=\frac{1}{8\pi^{2}}\int\limits _{\partial M}tr(A\wedge dA+\frac{2}{3}A\wedge A\wedge A)\:.\label{CS-invariante}
\end{equation}
Now we consider the gauge transformation $A\rightarrow g^{-1}Ag+g^{-1}dg$
and obtain
\[
CS(\partial M,g^{-1}Ag+g^{-1}dg)=CS(\partial M,A)+k
\]
with the winding number 
\[
k=\frac{1}{24\pi^{2}}\int\limits _{\partial M}(g^{-1}dg)^{3}\in\mathbb{Z}
\]
of the map $g:M\rightarrow G$. Thus the expression 
\[
CS(\partial M,A)\bmod1
\]
is an invariant, the Chern-Simons invariant. Now we will calculate
this invariant. For that purpose we consider the functional (\ref{CS-invariante})
and its first variation vanishes 
\[
\delta CS(\partial M,A)=0
\]
because of the topological invariance. Then one obtains the equation
\[
dA+A\wedge A=0\:,
\]
i.e. the extrema of the functional are the connections of vanishing
curvature. The set of these connections up to gauge transformations
is equal to the set of homomorphisms $\pi_{1}(\partial M)\rightarrow SU(2)$
up to conjugation. Thus the calculation of the Chern-Simons invariant
reduces to the representation theory of the fundamental group into
$SU(2)$. In \cite{FinSte:90} the authors define a further invariant
\[
\tau(\Sigma)=\min\left\{ CS(\alpha)|\:\alpha:\pi_{1}(\Sigma)\rightarrow SU(2)\right\} 
\]
for the 3-manifold $\Sigma$. This invariants fulfills the relation
\[
\tau(\Sigma)=\frac{1}{8\pi^{2}}\int\limits _{\Sigma\times\mathbb{R}}tr(F_{A}\wedge F_{A})
\]
which is the minimum of the Yang-Mills action 
\[
\left|\frac{1}{8\pi^{2}}\int\limits _{\Sigma\times\mathbb{R}}tr(F_{A}\wedge F_{A})\right|\leq\frac{1}{8\pi^{2}}\int\limits _{\Sigma\times\mathbb{R}}tr(F_{A}\wedge*F_{A})
\]
i.e. the solutions of the equation $F_{A}=\pm*F_{A}$. Thus the invariant
$\tau(\Sigma)$ of $\Sigma$ corresponds to the self-dual and anti-self-dual
solutions on $\Sigma\times\mathbb{R}$, respectively. Or the invariant
$\tau(\Sigma)$ is the Chern-Simons invariant for the Levi-Civita
connection.

\section{Chern-Simons invariant of Brieskorn spheres\label{sec:Chern-Simons-invariant-of-B}}

In \cite{FinSte:90} and\cite{KirKla:90,FreGom:91} an algorithm for
the calculation of the Chern-Simons invariant for the Brieskorn sphere
$\Sigma(p,q,r)$ is presented. According to that result, a representation
$\alpha:\pi_{1}(\Sigma(p,q,r)\rightarrow SU(2)$ is determined by
a tripel of 3 numbers $\left\langle k,l,m\right\rangle $ with $0<k<p,0<l<q,0<m<r$,
and the further relations
\begin{eqnarray*}
\frac{l}{q}+\frac{m}{r} & < & 1\qquad l\bmod2=m\bmod2\\
\frac{k}{p}+\frac{l}{q}+\frac{m}{r} & > & 1\\
\frac{k}{p}-\frac{l}{q}+\frac{m}{r} & < & 1\\
\frac{k}{p}+\frac{l}{q}-\frac{m}{r} & < & 1\:.
\end{eqnarray*}
 Then the Chern-Simons invariant is given by
\[
CS(\alpha)=\frac{e^{2}}{4\cdot p\cdot q\cdot r}\bmod1
\]
 with
\[
e=k\cdot q\cdot r+l\cdot p\cdot r+m\cdot p\cdot q\quad.
\]
 Now we consider the Poincaré sphere $P$ with $p=2,q=3,r=5$. Then
we obtain
\begin{eqnarray*}
\left\langle 1,1,1\right\rangle  &  & CS=\frac{1}{120}\\
\left\langle 1,1,3\right\rangle  &  & CS=\frac{49}{120}
\end{eqnarray*}
 and for the Brieskorn sphere $\Sigma(2,5,7)$
\begin{eqnarray*}
\left\langle 1,1,3\right\rangle  &  & CS=\frac{81}{280}\\
\left\langle 1,3,1\right\rangle  &  & CS=\frac{9}{280}\\
\left\langle 1,2,2\right\rangle  &  & CS=\frac{169}{280}\\
\left\langle 1,2,4\right\rangle  &  & CS=\frac{249}{280}\:.
\end{eqnarray*}
 In \cite{FinSte:90} the authors define a further invariant
\[
\tau(\Sigma)=\min\left\{ CS(\alpha)|\:\alpha:\pi_{1}(\Sigma)\rightarrow SU(2)\right\} 
\]
 for a homology 3-sphere $\Sigma$. For $P$ and $\Sigma(2,5,7)$
one obtains
\begin{equation}
\tau(P)=\frac{1}{120}\quad,\quad\tau(\Sigma(2,5,7))=\frac{9}{280}\label{eq:CS-for-B}
\end{equation}
and we are done.

\section*{References}

%\bibliographystyle{plain}
%\bibliography{DIFFBIB,foliation-gerbes,knots,inflation}

\end{document}